\documentclass[a4paper,twocolumn]{article}
\usepackage[dvips]{graphicx}

\begin{document}

\twocolumn[
\title{Growth Shape of Isotactic Polystyrene Crystals in Thin Films}
\author{Ken Taguchi \and Hideki Miyaji\and Kunihide Izumi\and 
Akitaka Hoshino \and Yoshihisa Miyamoto \and Ryohei Kokawa \\
Department of Physics, Graduate School of Science, \\
KyotoUniversity, Kyoto 606-8502 Japan \\
Department of Fundamental Sciences, Faculty of Integrated 
Human Studies, \\
Kyoto University, Kyoto 606-8501 Japan \\ 
Shimadzu Corporation, Kanagawa 259-1304 Japan \\}

\maketitle

\begin{abstract}
The crystal growth of isotactic polystyrene (it-PS) is
 investigated in very thin, 11 nm thick films. The it-PS crystals 
grown in the thin films show quite different morphology from that in 
the bulk. With decreasing crystallization temperature, the branching 
morphology in a diffusion field appears: dendrite and compact seaweed.
 The branching morphology is formed through a morphological 
instability caused by the gradient of film thickness around a crystal;
 the thicker the film thickness, the larger the lateral growth rate of
 crystals. Regardless of the morphological change, the growth rate as 
well as the lamellar thickness depends on crystallization temperature 
as predicted by the surface kinetics.
\end{abstract}
\vspace{3mm}


\vspace{1cm}
]

\section{Introduction}

Recently, crystal growth of polymers in thin films has attracted much 
attention on morphology and growth rate. Several authors have reported
 the appearance of diffusion controlled morphology in polymer thin 
films. Lovinger and Cais observed the single crystals crystallized 
from the melt of poly(trifluoroethylene) with branched morphology and 
discussed the morphology on the basis of the diffusion limited 
aggregation \cite{lovin}. Reiter and Sommer investigated the 
crystallization of poly(ethylene oxide) in quasi 2-dimensional 
monolayers and suggested that a diffusion field caused finger-like 
branched patterns observed \cite{reiter}; Sakai et al. reported the 
diffusion limited crystal growth of poly(ethylene terephthalate) in 
thin films \cite{imai}.

On the other hand, when isotactic polystyrene (it-PS) is crystallized 
from the melt in thin films down to 20 nm, the crystal has shown the 
same morphological change with crystallization temperature. Above 
$200^{\circ }$C the crystals grown are hexagonal plates, rounded 
hexagonal plates around $195^{\circ }$C, circular discs at $180^{\circ
 }$C (Fig.\ref{fig1}), and below $170^{\circ }$C two dimensional spherulites 
\cite{izumi1,sawamura}. Although the morphology is unchanged, the 
growth rates have been observed to decrease in thin films. The film 
thickness dependence of the growth rate was expressed by the following
 equation,

\begin{equation}
\frac{G(d)}{G(\infty)}=1-\frac{a}{d}\ ,
\label{gthin}
\end{equation}
where $G(d)$ is the growth rate in a film 
of thickness $d$, $G(\infty )$ is the growth rate in the bulk and $a$ 
is a constant of about 6 nm \cite{sawamura}. The constant $a$ was 
independent of crystallization temperatures, molecular weight and 
substrate material. 

In the present paper we report the morphology, growth rate and crystal
 thickness of it-PS crystals grown in ultrathin films, the thickness 
of which is less than the radius of gyration of a polystyrene molecule
 in the melt. In particular, the morphology will be discussed in the 
light of morphological instability in a diffusion field specific to 
thin polymer films, and compared with numerical simulations of crystal
 growth by Saito and Ueta \cite{saito} which took into account both 
surface kinetic process and diffusion process.

\section{Experimental}

The it-PS used in this study was purchased from Polymer Laboratory 
($M_{w}=590000,$ ${M_{w}}/{M_{n}}=3.4$, tacticity: 97\% isotactic 
triad). This molecular weight corresponds to 22 nm in radius of 
gyration of a polystyrene molecule in the melt. Ultrathin films of 
it-PS were prepared on a carbon-evaporated glass slide by spin coating
 at 4000 rpm; amorphous it-PS films with uniform thickness of about 11
 nm were obtained with use of a 0.4 wt\% cyclohexanone solution. The 
thickness of the film is determined by an atomic force microscope, AFM
 (SHIMADZU SPM-9500J). The films thereby obtained were crystallized 
isothermally at several temperatures, 180, 190, 195, 200, 205 and 
$210^{\circ }$C for a certain period of time in a hot stage (Mettler 
FP800). Before crystallization, the films were melted at $250^{\circ }
$C for 3 minutes, quenched to room temperature much lower than the 
glass transition temperature $T_{g}$ (=$90^{\circ }$C), and 
immediately elevated to a crystallization temperature. 
The lateral 
growth rate was determined by in-situ differential-contrast optical 
microscopy. Detailed morphology and structure of the crystals were 
investigated by the AFM and a transmission electron microscope, TEM 
(JEOL 1200EX II), at room temperature.

\section{Results}

Figure\ \ref{fig2} shows the AFM images of it-PS crystals grown at 
several crystallization temperatures in 11 nm thick films. It is to be
 noted that 11 nm is about a half of the radius of gyration of the 
molecules. The fact that these crystals can be observed by AFM 
necessitates that crystals are protruding from the surface of the 
surrounding amorphous it-PS. Indeed, in polymer crystallization in 
thin films, amorphous region close to the growth interface is always 
thinner by several nanometers than the region far from the interface. 
The surface of a lamellar crystal, including both the upper side and 
lateral growth front, is covered by an amorphous layer a few nm in 
thickness \cite{izumi1,izumi2,simon1,simon2}. A schematic view of the 
cross section is drawn in Fig.\ \ref{fig3}. At $210^{\circ }$C the 
morphology of the crystal is a hexagon with the {110} facets same as 
in the bulk. Below $210^{\circ }$C, however, the crystals were found 
to show morphology different from that in the bulk. At $205^{\circ }
$C, the {110} facets are no longer flat but have re-entrant corners to
 form a star-like shape. Side branches begin to appear at $200^{\circ }
$C and appears a dendrite similar to a snowflake at $195^{\circ }$C.
At $190^{\circ }$C, a dendrite has a number of side branches and the 
crystal still has 6-fold symmetry. Below $180^{\circ }$C, we can 
observe the splitting of growing tips and many irregular branches to 
form the morphology similar to the compact seaweed (CS) \cite{brener} 
or dense branching morphology (DBM) \cite{dbm} with a circular and 
convex envelope. The mean width of branches, $w$, measured directly 
from AFM images, decreases with decreasing crystallization temperature
 as shown in Fig.\ \ref{fig4}. The electron diffraction pattern from
 a crystal grown at $180^{\circ }$C is shown in Fig.\ \ref{fig5} with 
its bright field image. Although there are many irregular branches, 
the pattern reveals that this crystal has a single crystallographic 
orientation with the chain axis perpendicular to the film surface. In 
fact, all the crystals in Fig.\ \ref{fig2} were single crystals. With 
decreasing crystallization temperature, the morphology of the it-PS 
single crystals in ultrathin films therefore changes from a 
nucleation-controlled faceted hexagonal plate to the branching 
morphology observed in the crystal growth in a diffusion field.

\begin{figure}
 \begin{center}
  \includegraphics[width=8cm,clip]{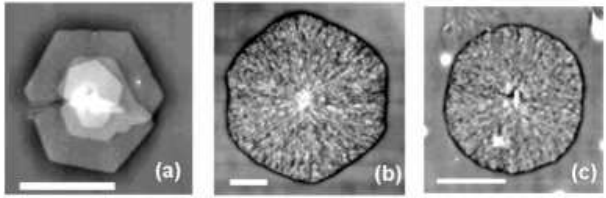}
 \end{center}
\caption{AFM images of it-PS crystals grown in films thicker than $50nm$ at 
(a) $210^{\circ }$C for 2hr45min, (b) 195$^{\circ }$C, 1hr (c) 180$^{\circ }$C,
 20min. Scale bars represent 5$\mu$m. These crystals consist of not single
 lamella but many overgrowth lamellae. The morphology at each temperature is
 the same as in the bulk.}
\label{fig1}
 \end{figure}

\begin{figure}
 \begin{center}
  \includegraphics[width=7cm,clip]{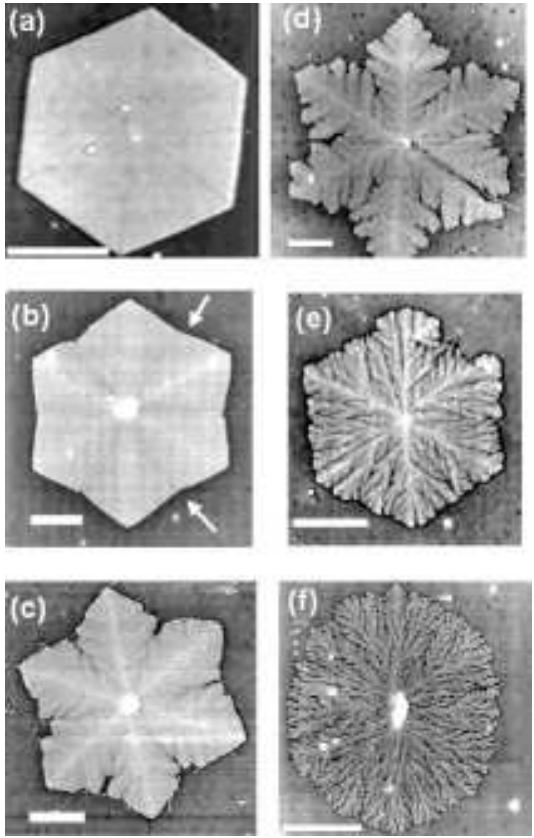}
 \end{center}
\caption{AFM images of it-PS crystals grown in 11nm-thick film (a) at 
$210^{\circ }$C for 4 hr, (b) $205^{\circ }$C, 4 hr, (c) $200^{\circ }
$C, 4 hr, (d)$195^{\circ }$C, 3 hr, (e) $190^{\circ }$C, 1 hr, (f) 
$180^{\circ }$C, 1 hr. Scale bars are 5$\mu$m. The arrows in (b) show 
the re-entrant sites, where the growth rates decrease during growth.}
\label{fig2} 
\end{figure} 

\begin{figure}
 \begin{center}
  \includegraphics[width=8cm,clip]{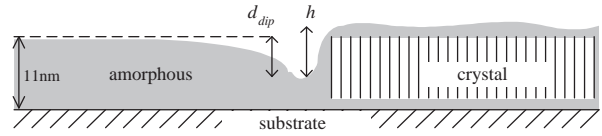}
 \end{center}
\caption{A schematic view of the cross section. $h$ is height 
difference between the upper surface of the crystals and the bottom of the 
concave amorphous surface near the growth face, $d_{dip}$ is the depth of the
 concave region. The vertical scale is magnified about 100 times 
compared with the horizontal scale.}
\label{fig3} 
\end{figure} 

\begin{figure}   
 \begin{center}
  \includegraphics[width=8cm,clip]{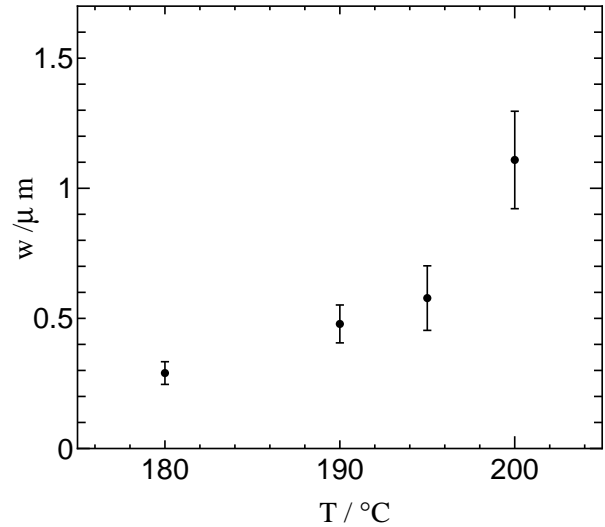}
 \end{center}
\caption{The mean width of branches $w$ vs. crystallization 
temperature $T$ in it-PS 11nm-thick films.}
\label{fig4} 
\end{figure}

\begin{figure}   
 \begin{center}
  \includegraphics[width=6.5cm,clip]{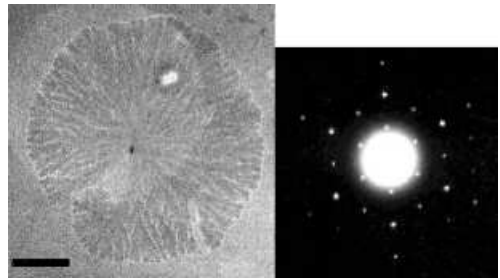}
 \end{center}
\caption{Bright field image and electron diffraction pattern of an 
it-PS crystal grown at $180^{\circ }$C in ultrathin film. Scale bar is
 2$\mu$ m}
\label{fig5} 
\end{figure}

\begin{figure}   
 \begin{center}
  \includegraphics[width=7cm,clip]{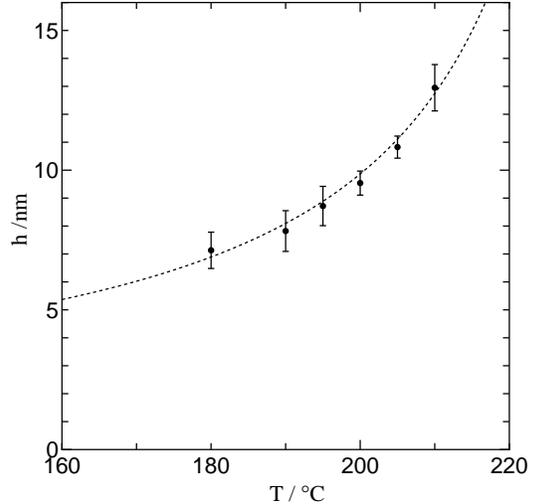}
 \end{center}

\caption{The height differeence $h$ in Fig.2 vs. crystallization 
temperature $T$. The dotted curve shows the theoretical temperature 
dependence of the lamellar thickness $l_{g}$ [12]: $l_{g}^{*}=[2\sigma
 _{e}T_{m}^{0}/\Delta h_{f}(\Delta T)]+\delta l$; $\delta 
l=\frac{kT}{2b\sigma}\cdot \frac{(4\sigma /a)+\Delta f}{(2\sigma 
/a)+\Delta f}$, $\sigma =7.65ergs/cm^{2}$,$\sigma _{e}=34.8ergs/cm^{2}$, 
$\Delta h_{f}=9.40\times 10^{8}ergs/cm^{3}$, $T^{0}_{m}=242^{\circ }C$, 
$a=6.3\AA $, $b=10.9\AA $}
\label{fig6} 
\end{figure}

\begin{figure}   
 \begin{center}
  \includegraphics[width=7cm,clip]{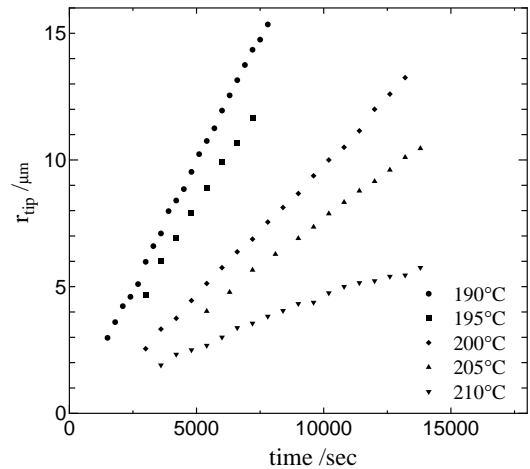}
 \end{center}
\caption{Time developments of $r_{tip}$ of an it-PS crystal in 
11nm-thick films at each crystallization temperature.}
\label{fig7} 
\end{figure}

\begin{figure}   
 \begin{center}
  \includegraphics[width=7cm,clip]{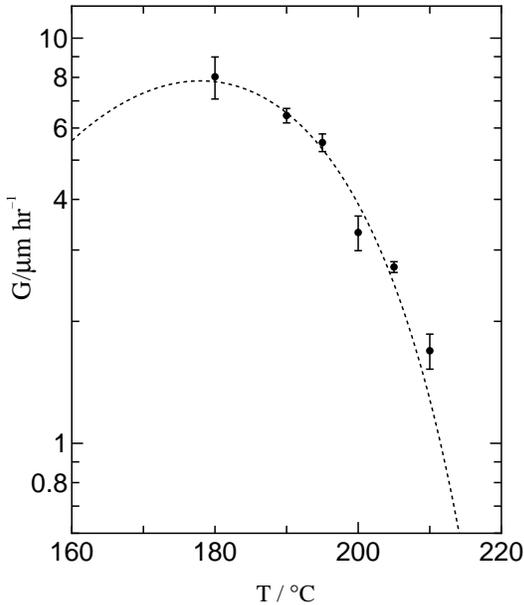}
 \end{center}
\caption{Growth rates vs. crystallization temperatures in 11nm-thick 
films. The dotted curve shows the growth rate in the bulk multiplied 
by a factor 5/11.}
\label{fig8} 
\end{figure}

In polymer crystallization, important aspects besides morphology are 
the thickness of a lamellar crystal and the lateral growth rate 
\cite{hoffman}. The thickness of each crystal in ultrathin films is 
almost constant during growth as revealed by the uniform contrast in 
the AFM image of crystal surfaces (Fig.\ \ref{fig2}). Since 
amorphous layers cover the upper and lateral surfaces of crystals and 
should also exist between the crystals and the substrate 
\cite{simon1}, it is difficult to determine the absolute value of the 
lamellar thickness. However, as shown in Fig.\ \ref{fig3}, the height 
difference $h$ may be a relative 
measure of the thickness since the amorphous layers covering the crystals
 are considerably thin compared with lamellar thickness and the depth of 
concave region, $d_{dip}$, has an almost constant value of several nm in 
ultrathin films.
The height differences measured increased with crystallization 
temperature, agreeing well with theoretical curve of lamellar thickness 
\cite{hoffman} as shown in Fig.\ \ref{fig6}.
 This result suggests that crystallization temperature dependence of 
the lamellar thickness in ultrathin films is almost the same as in the
 bulk \cite{miyamoto1}. For the growth rate determined by the time 
development of the separation of the farthest tip $r_{tip}$ from the 
center of the crystal (Fig.\ \ref{fig7}), the rate was found to be 
constant during growth (linear growth) and was about $5/11$ of those 
in the bulk at each crystallization temperature (Fig.\ 
\ref{fig8}), agreeing with Eq. (\ref{gthin}) for $d =11$ nm and 
$a = 6$ nm. As for the re-entrant site of the crystal between the 
farthest tips shown by the arrows in Fig.\ \ref{fig2} (b), on the other hand,
 the grwoth rate was not constant but decreasesd during growth, and the
 film thickness 
in front of the re-entrant site decreases with time. 
It should be noted, however, that the lamellar crystal thickness is almost
 constant throughout the crystal including the re-entrant site with 
different growth rate.
Consequently, the lamellar thickness and the linear growth rate at the
 growing tip in the ultrathin films depend on crystallization 
temperature as in the bulk. Hence the surface kinetics 
\cite{hoffman,sg,doye} of the crystallization in ultrathin films is 
the same as that in the bulk regardless of the drastic morphological 
change.

\section{Discussion}

According to the theory of crystal growth, dendrites and CS can appear
 in the crystal growth in a diffusion field.
When a part of the flat interface grows faster than the other by some 
fluctuation, the advanced part can grow faster owing to a gradient 
(e.g. of concentration) in the diffusion field, 
and the deformation is enhanced. Hence the flat interface becomes unstable, 
resulting in dendrites or CS. The morphology of it-PS 
crystals grown in ultrathin films below $205^{\circ }$C therefore 
suggests that a diffusion process plays an important role in the 
crystal growth. In fact, the morphological change with increasing 
supercooling or driving forces similar to the present study has been 
reported in both experiment and simulation on crystal growth in a 
diffusion field \cite{saito,ovsi,oswald,kobaya,yokoyama}. For example,
 in the growth of snow crystals, both a diffusion process and a 
surface kinetic process are closely reflected in the variety of growth
 shape 
\cite{yokoyama}. Saito and Ueta performed the Monte Calro (MC) 
simulation on the crystal growth in a diffusion field \cite{saito} and
 found the morphological transition from a polygon through a dendrite 
to CS with increasing driving force, which is very similar to our 
result. A further important point in their study is that those 
crystals show a linear growth and surface kinetics controls the growth
 until morphological transition to CS occurs. Hence in the 
crystallization of it-PS in ultrathin films, the surface kinetics 
keeps working at the growth tip while the morphology is controlled by 
diffusion process.

Here we must ask a question: what kind of diffusion field causes the 
instability in the present case? The diffusion field is generally 
either that of concentration or temperature with a gradient at growth 
face; the gradient causes the instability of flat growth faces to give
 rise to dendrites or DBM depending on supersaturation or supercooling
 and the strength of anisotropy in surface free energy. However, in 
the crystallization of polymers, the growth rate is so slow that the 
thermal diffusion length is much larger than crystal size, and hence 
no appreciable temperature gradient exists \cite{toda}; we can neglect
 the effect of thermal diffusion. In the melt growth, no concentration
 gradient exists other than that of any kind of impurity. However, we 
need not take into consideration even the effect of impurity diffusion
 because no morphological instability is observed on the crystal 
growth in the thicker films in spite of much higher growth rates. 

Detailed inspection of the AFM images in Fig.\ \ref{fig1} elucidates 
the gradient of thickness of melt around a crystal; the thickness of 
melt increases gradually with separation from the crystal edge to 
surrounding melt.
 The gradient is formed owing to the protrusion of a crystal from the 
surrounding melt \cite{izumi2}. This gradient makes self-diffusion 
field in front of growing crystals. Since the thickness in thin films 
is proportional to the amount of polymer segment per unit area, the 
thickness of melt, $d$, can correspond to the concentration; we can 
recognize this morphological instability on the analogy of that of the
 crystal growth from solutions. Therefore it must be this gradient 
of thickness around the crystal that introduces the instability.
 The effect of this gradient on the growth rate may be related with the
 film thickness dependence of the growth rate (eq. (\ref{gthin})). However,
 the detail of this relation is yet to be discussed. The film 
thickness in the previous study \cite{sawamura} was above 20 nm and 
much thicker than the depletion of the thickness around a crystal, 
several nm, and hence the effect of the gradient of thickness on the 
growth rate was so small that no morphological instability was 
observed.

Lastly, we consider one of the main structures of branching 
morphology, the width of branches $w$ (Fig.\ \ref{fig4}). The 
characteristic length of diffusion-controlled growth is generally 
given by the stability length $\lambda _{s}=2\pi \sqrt{d_{0}l_{D}}$, 
where $d_{0}$ is a capillary length and $l_D$ a diffusion length 
\cite{ms,kp}. Indeed, the occurrence of spherulites, which consist of 
stacking elongated lamellae, was accounted for with the 
diffusion-controlled growth, and their characteristic dimension was 
predicted to be scaled with $\lambda _{s}$\cite{golden}. In the MC 
simulation mentioned above, on the other hand, the structure of 
dendrites is characterized by the diffusion length $l_D$. In either 
case, the change in $l_D$ with decreasing temperature allows us to 
predict the qualitative change in characteristic length. The diffusion
 length is given by $l_D=2D/G$, where $D$ is the diffusion coefficient
 and $G$ is the growth rate; in the present experiment of melt 
crystallization, we assume that $D$ represents the self-diffusion 
coefficient of polymer chains. In the range of crystallization 
temperature investigated the growth rate increases rapidly with 
decreasing crystallization temperature (Fig.\ \ref{fig7}), while the 
diffusion coefficient generally decreases with decreasing temperature,
 and hence $l_D$ decreases rapidly with decreasing crystallization 
temperature, agreeing with the experimental result (Fig.\ \ref{fig4}).

\section{Conclusions}

We have clearly shown that the morphology of it-PS crystals grown in 
ultrathin films changes from the faceted morphology to that in the 
diffusion field; the lateral growth interface becomes unstable with 
increasing supercooling. This morphological instability is considered 
to result from the gradient of film thickness around the crystal; the 
formation of gradient is specific in polymer crystallization in thin 
films. The growth rates and the lamellar thickness show that the 
surface kinetics keeps working on the crystallization in ultrathin 
films while the morphology is controlled by diffusion process. 

\section*{Acknowledegments}

This work was supported partly by Grant-in-Aid for Science Research on
 Priority Areas, "Mechanism of Polymer Crystallization" (No12127204) and
Grant-in Aid for Exploratory Research No12874048 
from The Ministry of Education, Science, Sports and Culture.

\end{document}